\documentclass[12pt,preprint]{aastex}

\newcommand{\myemail}{levenson@physics.ucla.edu}

\shorttitle{Spitzer over DIRBE}
\shortauthors{Levenson \& Wright}

\begin{document}

\title{Probing the 3.6 Micron CIRB with \it Spitzer \bf in 3 DIRBE Dark Spots}

\author{L. R. Levenson \& E. L. Wright }
\affil{Department of Physics and Astronomy, University of California, Los Angeles, CA 90095-1562}
\email{\myemail}

\begin{abstract}
 We observed three regions of the sky with \it Spitzer \rm in which the Cosmic InfraRed Background (CIRB) has been determined at 3.5 $\mu$m using the method of subtracting 2MASS stellar fluxes from zodiacal light subtracted DIRBE maps.  For each of these regions we have obtained 270 seconds of integration time per pixel with IRAC on \it Spitzer \rm over the central square degree.  We present galaxy counts in each of these approximately 1 square degree IRAC surveys.  Along with deep galaxy counts in the Extended Groth Strip and GOODS North, we are able to compare the galactic contribution to the CIRB with the ``DIRBE minus 2MASS'' determined L-band CIRB.  Using the profile-fit photometry package GIM2D, we find a substantially larger flux contribution to the CIRB than that determined using aperture photometry.  We have also made the first rigorous analysis of the uncertainties in determining the CIRB via galaxy counts in \it Spitzer \rm images using a Monte Carlo Markov Chain simulation of our data analysis.  Using a simple broken power law model for galaxy counts as a function of magnitude we find a most probable contribution to the CIRB from galaxies at 3.6 $\mu$m of $10.8^{+2.1}_{-1.1}$ kJy sr$^{-1}$ ($9.0^{+1.7}_{-0.9}$ nW m$^{-2}$ sr$^{-1}$).  Even with this restricted model, however, we find that galaxy counting does not strongly constrain the CIRB from above. We are able to find solutions in which the CIRB runs away to large intensities without the need for an additional diffuse source.

\end{abstract}


\section{Introduction}

The Cosmic InfraRed Background (CIRB) is the aggregate of the short wavelength radiation from the era of structure formation following the decoupling of matter and radiation in the early universe, redshifted to near infrared wavelengths by cosmological redshifting and far infrared wavelengths by dust-reprocessing, i.e. absorption and re-emission of starlight by intervening dust. Particle decay models also allow for a contribution to the CIRB from photons produced in the decay of weakly interacting massive particles such as big bang relic neutrinos \citep{bond86}.  Hence, the cumulative light from galaxies is a strict lower limit on the CIRB.  A determination of the total contribution of resolved galaxies to the CIRB was made recently via galaxy number counts by \citet{faz04}.  Using the InfraRed Array Camera (IRAC) on the Spitzer Space Telescope  \citep{eis04}, surveys were done of the Bo\"{o}tes region, the Extended Groth Strip and a deep image surrounding the QSO HS 1700+6416.  After integrating the light from galaxies from the 10th to the 21st magnitudes, a total integrated intensity of 6.5 kJy sr$^{-1}$ (5.4 nW m$^{-2}$ sr$^{-1}$) at 3.6 $\mu$m is reported. This is less than half of the CIRB determined at that wavelength by \citet{dwe98}, \citet{gor00}, \citet{wrr00}, \citet{wrj01} and \citet{me07a} which directly measured the CIRB via the subtraction of foregrounds from all sky maps from the Diffuse InfraRed Background Experiment (DIRBE) on the COsmic Background Explorer (COBE, see \citet{bog92}).  \citet{me07a} determine a 3.5 $\mu$m CIRB of 15.6 $\pm$ 3.3 kJy sr$^{-1}$ (13.3 $\pm$ 2.8 nW m$^{-2}$ sr$^{-1}$).  This leaves a greater than 2$\sigma$ gap between the direct measurements of the CIRB and the lower limit from galaxy counts.  To investigate this discrepancy, we have obtained \it Spitzer \rm observations of the central square degree of three of the ``dark spots'' used in the ``DIRBE Minus 2MASS'' direct determinations of the CIRB, where ``DIRBE Minus 2MASS'' refers to the series of papers by \citet{elw01}, \citet{wrj01} and \citet{me07a} which estimated the CIRB intensity by subtracting the \citet{wr98} zodiacal light model, 2MASS stellar fluxes and a faint star count model from the total DIRBE intensity.  The galaxy flux contribution to the CIRB determined by \citet{faz04} uses Bertin's \citep{ber96} Source Extractor (SE) for both source detection and aperture photometry.  While we also use SE for source detection, we employ a profile-fit photometry method using the IRAF package GIM2D \citep{sim01} in an attempt to include more of the faint fringes of faint galaxies in our cumulative count of the light from these galaxies.  

To this point, only lower limits from galaxy counts have been reported in the literature.  Here we have estimated uncertainties resulting from counting (Poisson) and cosmic variance, as well as from the flux measurements, whether from profile-fit or aperture photometry.  At faint magnitudes ($m > 16$) the sky becomes so crowded that confusion becomes a significant barrier to flux estimation using either method.  Errors become more pronounced at faint magnitudes where confusion is significant, but even at bright magnitudes profile fitting tends to overestimate the flux, while aperture photometry underestimates it.  We have corrected for these errors using a rigorous analysis employing a  Monte Carlo Markov Chain (MCMC) simulation of our analysis by constructing a simple broken power law model for the ``real'' galaxy counts and using the Markov Chain to explore the parameter space of models that could lead to our data.  Two simple priors were used, one requiring that the integrated flux converge and another weighting the bright end galaxy counts near Euclidean counts.  We are now able, drawing from the posterior distribution, to compute the CIRB intensity with the first robust error bars on the galactic contribution to the CIRB.

\section{Data}

Our most recent measurement of the L-band CIRB \citep{me07a} used the method of \citet{elw01} to subtract stellar fluxes from the 2 Micron All Sky Survey (2MASS) Point Source Catalog (PSC) from zodiacal light subtracted DIRBE maps in 40 regions of the sky.  All 40 of these regions are at high galactic latitude and scattered widely in ecliptic latitude.  The three regions of the sky chosen for observation with \it Spitzer \rm were selected from these 40 DIRBE ``dark spots.'' The regions chosen are centered at widely different ecliptic latitudes of +20 (region 2), +50 (region 1) and +70 (region 3) degrees.  Actual region coordinates are shown in Table~\ref{coords}.  The observing strategy was to use the mapping capabilities of IRAC to map the central square degree of each of these regions with a total integration time of 270 seconds per pixel.  Observations were divided into 3 sets of 3-30 second frames at each pointing, always at a new point in the dither table.  Thus, each pointing was visited on nine separate occasions with three observations approximately 45 seconds apart as the dithered map was executed within each AOR, and returning three times to repeat this process anywhere from hours to days to months later depending on the scheduling of separate AOR's for observation.  This redundancy allowed robust removal of artifacts such as cosmic rays, whie the dithering ensured that a bad pixel would be backed up with several observations with good pixels.  Once all of the individual observations were mosaiced, the total usable coverage obtained in each region is 0.60, 0.84 and 0.82 square degrees in regions 1, 2, and 3 respectively, where region 1 is smaller due to a long gap between observations that allowed a rotation of the telescope. 

\begin{figure}
\epsscale{.6}
\plotone{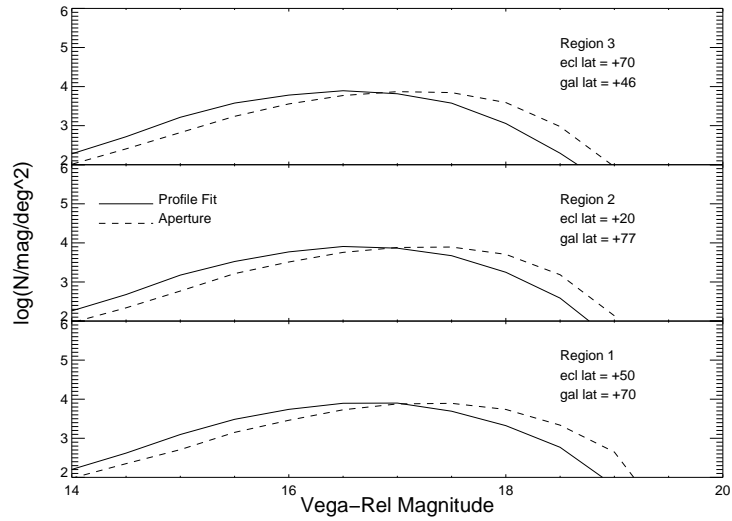}
\caption{Galaxy Counts; Aperture vs. Profile-Fit: Galaxy counts vs. magnitude using both SExtractor's mag\_auto aperture photometry (dashed) and GIM2D profile fit photometry (solid).  Profile fit magnitudes are approximately 0.5 mag brighter on average than aperture magnitudes for the same sources. } 
\label{aper}
\end{figure}

As shown in Figure~\ref{aper}, source counts in these three regions begin to be severely incomplete between the 16th and 17th magnitudes. Galaxies at these magnitudes are the peak contributors to the CIRB.  In order to obtain counts at magnitudes deeper than the peak contribution to the CIRB, the use of deeper images was required.  Counts at intermediate depths were done using a deep image of the Extended Groth Strip (EGS) observed as part of the GTO project ``The IRAC Deep Survey.''  The deepest counts were done using the Great Observatories Origins Deep Survey (GOODS) North field \citep{dic02}.     

\section{Analysis}
\label{Analysis}

\begin{figure}
\epsscale{.6}
\plotone{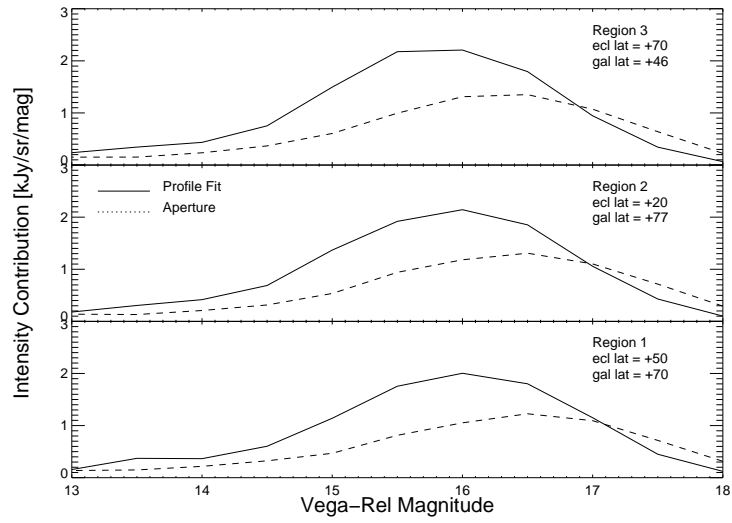}
\caption{Galaxy Intensity; Aperture vs. Profile-Fit: The Intensity contribution to the CIRB from the above galaxy counts is $\approx$ 1.6 times brighter when profile fit magnitudes are used.} 
\label{aper2}
\end{figure}

Basic Calibrated Data (BCD) images were combined into mosaics in each region using MOPEX.  Source detection in each region, as well as in the deep image mosaics, was then performed using Source Extractor (SE) \citep{ber96}.  A conservative detection limit required 5 adjacent pixels at least 5$\sigma$ above the SE determined background.  Aperture magnitudes for each source were also determined using SE's mag\_auto which uses an adjustable elliptical \citep{kro80} aperture.  However, since aperture photometry can miss the faint fringes of galaxies, especially in crowded fields where large apertures can not be used, a more sophisticated method of profile fit photometry was employed. The IRAF package GIM2D \citep{sim01} fits a 12-parameter, bulge plus disk profile to each source and determines the photometry from the resulting model.  A comparison of the aperture and profile-fit galaxy-only counts in each region are shown in Figure~\ref{aper}. Removal of stars from the total counts is described in detail below.  The resulting intensity contribution of these galaxies to the CIRB using the two methods are shown in Figure~\ref{aper2}.  Profile fitting results in magnitudes approximately 0.5 mag brighter, on average, for the same source.  This results in approximately 1.6 times more intensity contributed to the CIRB.  All magnitudes quoted here are Vega-relative magnitudes where Vega has a flux density at 3.6 $\mu$m of 277.5 Jy \citep{faz04}.

In order to determine the contribution of galaxies to the CIRB, stars must be removed from the raw source catalogs.  At brighter magnitudes, stars and galaxies can be separated morphologically.  GIM2D returns a half-light radius for each fitted source.  This half-light radius is as reliable an indicator of a galaxy as visual inspection of the galaxy image, for galaxies brighter than magnitude 13.5.  At magnitudes fainter than 13.5, both visual inspection of the source image and the GIM2D half-light radius become ambiguous indicators of morphology and another method of star/galaxy separation is required.  Regions 1 \& 2, as well as the EGS and GOODS-N are covered by the Sloan Digital Sky Survey (SDSS) data release 5 \citep{adel07}.  All of the detected sources in the resulting SE/GIM2D catalogs from these images were fed into the SDSS/Skyserver Imaging CrossID web tool.  Approximately half of the sources in our IRAC images had matches within 3" in the SDSS catalog.  Any source ($m > 13.5$) with a match within 3" classified by SDSS as a star was removed from the catalog.  All sources classified by SDSS as galaxies, along with all sources red enough to be detected in our IRAC images and not detected by SDSS were considered to be galaxies and included in the total intensity of the CIRB.  The nearby 8th magnitude galaxy NGC 5927 in region 1 was not included in the total CIRB estimate.

\begin{figure}
\epsscale{.80}
\plotone{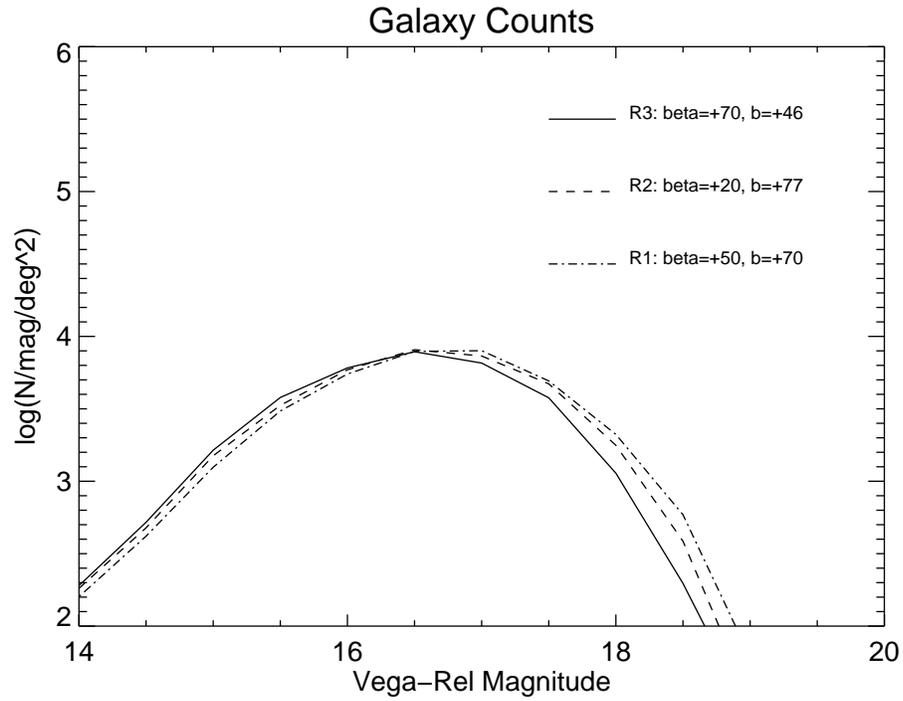}
\caption{Profile fit counts vs. magnitude in regions 1 2 \& 3 (R1, R2, R3) where beta is the ecliptic latitude and b is the galactic latitude of the region center.  Fractional variance in the total counts per steradian between the three regions is less than 1\%. The fact that the total galaxy counts do not show a dependence on galactic latitude, b, indicates successful removal of stellar sources in these three regions and supports the above assumption of a csc$|$b$|$ scaling for the star counts.} 
\label{gal123}
\end{figure}

As region 3 was not covered by SDSS, it was necessary to remove stars with $m > 13.5$ statistically using the average of the star/galaxy ratios at each magnitude in regions 1 and 2 multiplied by the ratio of the sines of the galactic latitudes.  This assumes a csc$|$b$|$ scaling for star counts.  These ratios were applied to the raw source counts at each magnitude in region 3 using, 
\begin{equation}
\left(\frac{n_{star}}{n_{gal}}\right)_3 =  0.5 \left[\left(\frac{n_{star}}{n_{gal}}\right)_{2}\left(\frac{\sin{b_2}}{\sin{b_3}}\right) + \left(\frac{n_{star}}{n_{gal}}\right)_{1}\left(\frac{\sin{b_1}}{\sin{b_3}}\right)\right]   \, ,
\end{equation}
where $b_i$ is the galactic latitude of region $i$.  Resulting galaxy-only counts in all three regions are shown in Figure~\ref{gal123}.  Hereafter, all counts refer to those determined using profile-fit magnitudes.  While the fractional variance $\delta N/\langle N \rangle$ in the galaxy counts per unit solid angle at a particular magnitude is as high as 11\%,  $\delta N/\langle N \rangle$ of the \it total counts \rm per unit solid angle between the three regions is less than 1\%. The fact that the total galaxy counts do not show a dependence on galactic latitude, b, indicates successful removal of stellar sources in these three regions and supports the above assumption of a csc$|$b$|$ scaling for the star counts.

\begin{figure}
\epsscale{.80}
\plotone{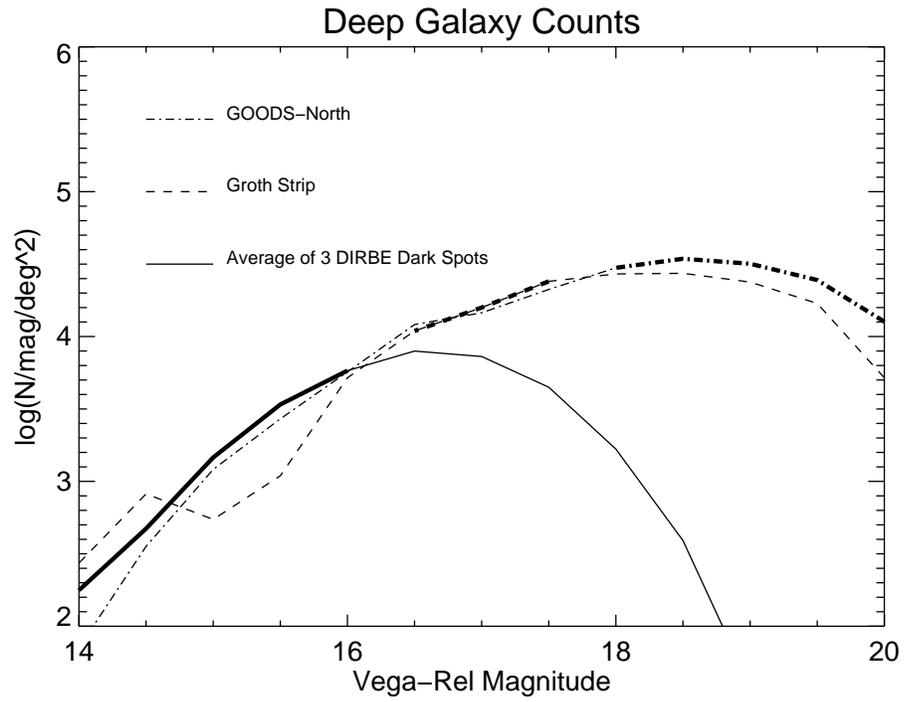}
\caption{Averaged profile-fit galaxy counts from the three regions along with deep counts from the EGS and GOODS North fields using the same source detection and photometry.  Boldness indicates the magnitude range over which each data set was used in compiling the total observed galaxy counts.} 
\label{deepgal}
\end{figure}

Figures~\ref{gal123} and \ref{deepgal} clearly show that counts in regions 1-3 are incomplete by $m = 16.5$.  Since this is near the peak contribution to the CIRB, accounting for this incompleteness is crucial to obtaining an estimate of the galactic contribution to the CIRB.  We chose to use deeper images to obtain real counts at these important magnitudes rather than rely on an estimate of the incompleteness in our surveys.  Figure~\ref{deepgal} shows the average of the counts in our three regions along with the deeper counts, obtained using the exact same methods, from the EGS and GOODS-N. Final observed counts are then the averaged counts from our three regions for $m \leq$ 16.0, averaged counts from the EGS and GOODS-North for 16.5 $\leq m \leq$ 17.5, and GOODS-North for $m \geq$ 18.0.

Completeness was estimated using the standard Monte Carlo method of inserting sample galaxies of known magnitudes into the science images at known coordinates and re-running the source detection.  The sample galaxy images used were model images created by GIM2D in the process of fitting actual galaxies.  Sources were considered recovered if a new source was detected within $\pm$ 1.5 pixels of the insertion coordinates, or, if an old source was brightened by at least half of the inserted galaxies flux.  In the shallow fields, for magnitudes less than 16.5, at each half magnitude, 200 copies of at least 31 different model galaxies were inserted and re-detection attempted in 4 iterations making a minimum of 24800 sample sources inserted.  At fainter magnitudes, since the deep fields are smaller and more crowded, only 40 copies of each galaxy were inserted in a single iteration.  However, at least 34 different model galaxies were used and 10 iterations of the process were performed resulting in a minimum of 13600 sample sources inserted at each half magnitude down to $m = 19.5$.  Completeness is already below 40\% at $m = 19.5$, so observed counts were cut off at this magnitude.  

Completeness estimates for magnitudes 10.5 - 19.5 are shown in Table~\ref{uncandcomp}.  Our analysis finds the deep images to be substantially less complete than the \citet{faz04} estimates, even in the same image (EGS).  Several effects may be causing this difference.  The first is the detection limit used.  While SE was used for source detection in both cases, \citet{faz04} use a 2.5$\sigma$ detection limit while we have used a more conservative 5$\sigma$ limit.  Thus, given the same spatial scale, they will detect a galaxy with a surface brightness twice as faint, and thus go 0.75 magnitudes deeper.  In addition, the magnitude at which the completeness is determined is given here by our profile-fit magnitudes as opposed to their aperture magnitudes.  We have shown here that our magnitudes are, on average, approximately 0.5 magnitude brighter for the same source than those determined with SE aperture photometry.  Deep field photometry in \citet{faz04} was done with custom software, however, so it is difficult to know just how these compare.  

\begin{figure}
\epsscale{.80}
\plotone{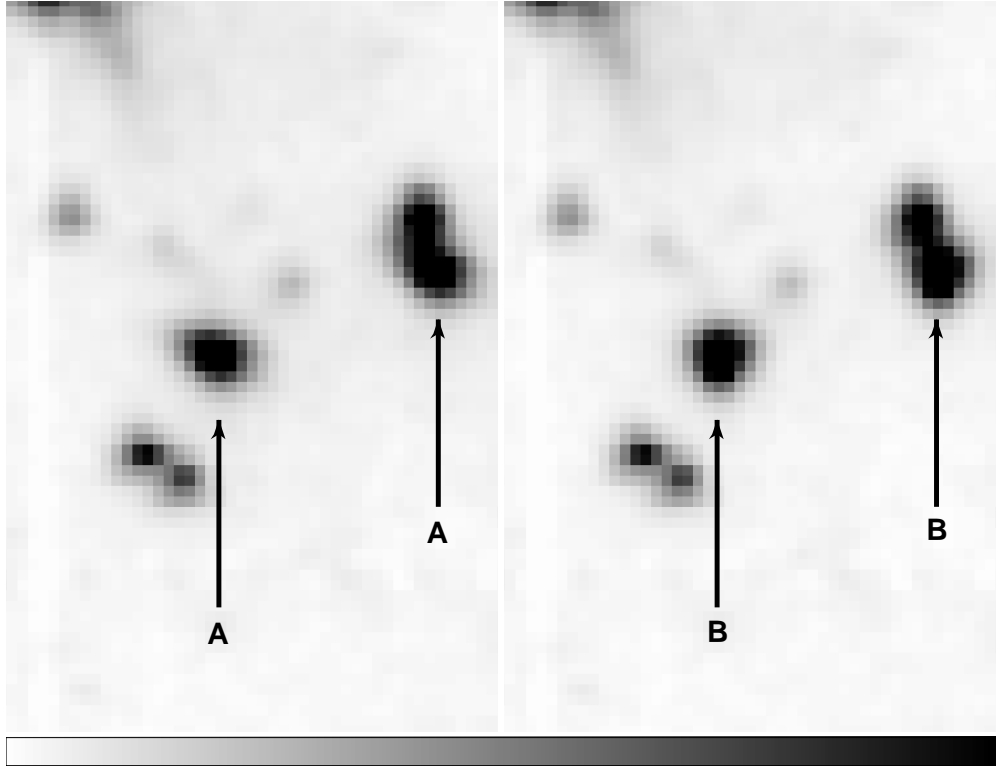}
\caption{The left hand frame shows two of the 40 copies of galaxy A that were inserted into the science image for completeness testing.  Galaxy B was inserted at the same 40 randomly generated coordinates.  Both galaxies have profile-fit magnitudes of $m = 16.5$.  However, galaxy A has a peak surface brightness of 207.0 kJy sr$^{-1}$ and is only detected 60\% of the time while galaxy B has a peak surface brightness of 259.2 kJy sr$^{-1}$ and is detected 88\% of the time.} 
\label{lsbhsb}
\end{figure}

\begin{figure}
\epsscale{.80}
\plotone{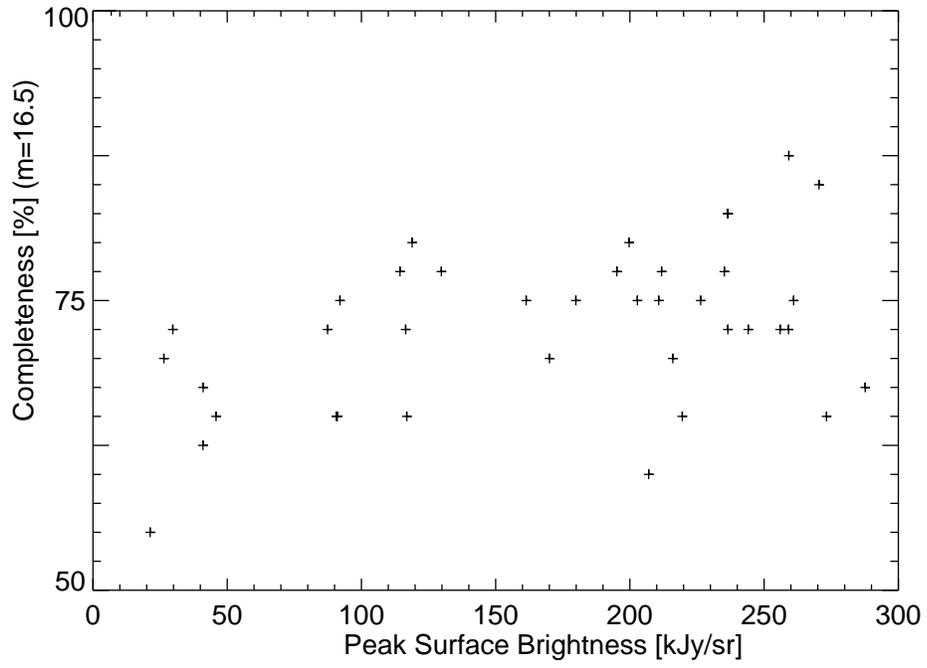}
\caption{Completeness vs. peak surface brightness for galaxies at m = 16.5.} 
\label{compvspeak}
\end{figure}

\begin{figure}
\epsscale{.80}
\plotone{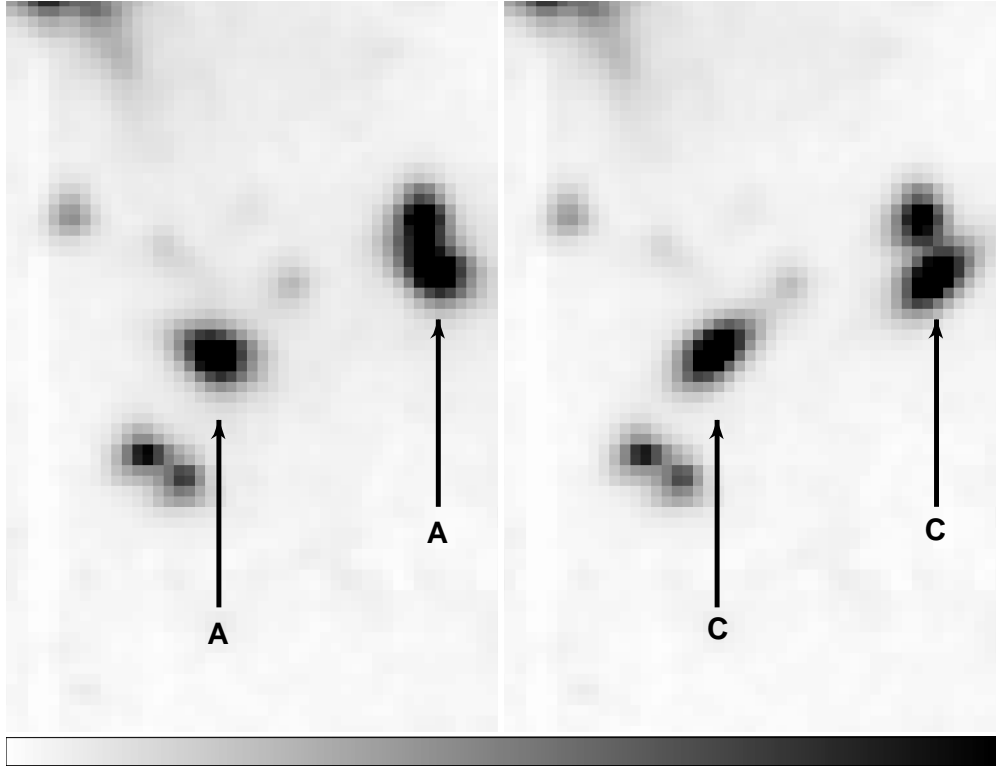}
\caption{The inserted copies of galaxy A in the left hand frame are the same as those in the left panel of Figure~\ref{lsbhsb}.  Here, the right hand frame shows galaxy C which has a similar peak surface brightness to galaxy A (211.9 and 207.0 kJy sr$^{-1}$ respectively) and both still have profile-fit magnitudes of $m = 16.5$.  However, galaxy C has a higher ellipticity and smaller half-light radius and while galaxy A is detected 60\% of the time, galaxy C is detected 78\% of the time.} 
\label{morph}
\end{figure}

The remaining difference may be due to the observational characteristics of the particular galaxies used for completeness estimation.  Here, the galaxies inserted into the science images were chosen at random from each half magnitude bin.  This results in a wide variety of observational characteristics such as the peak of the surface brightness distribution, the ellipticity, bulge fraction, and spatial extent, all of which affect the rate at which galaxies of the same magnitude are recovered.   For example, Figure~\ref{lsbhsb} shows a sample of two of the galaxies used, labeled A and B, which have very different peak surface brightnesses.  The indicated galaxies in both the right and left panels were inserted at the same coordinates and both have profile-fit magnitudes of $m = 16.5$.  However, the galaxy inserted into the left frame (galaxy A) has a peak surface brightness of 207.0 kJy sr$^{-1}$ while the galaxy inserted into the right hand frame (galaxy B) peaks much higher, at 259.2 kJy sr$^{-1}$.   And while galaxy A actually has a larger total flux,  it is detected at only 24 of 40 inserted positions in a single iteration compared with  35 of the 40 insertions for galaxy B.  However, the peak of the surface brightness distribution is not the only feature affecting completeness.  Figure~\ref{compvspeak} shows that while there is a trend with the peak surface brightness, the completeness at a given peak surface brightness is also widely scattered.  Figure~\ref{morph} shows that the way the remaining flux is distributed about the peak, affected by the galaxy's inclination and morphology, also plays a role.  The left hand frame is the same as that in Figure~\ref{lsbhsb}, where galaxy A has a peak surface brightness of 207.0 kJy sr$^{-1}$ and a 60\% detection rate.  The galaxy inserted into the right hand frame (galaxy C) has a similar peak surface brightness of 211.9 kJy sr$^{-1}$, but a detection rate of 78\%.    Visually, it is obvious that galaxy C has a higher ellipticity than galaxy A (0.7 vs. 0.4, based on the GIM2D fit).  Less obvious to the eye, however, is that the latter has a half-light radius approximately 40\% larger.  For this work it is not necessary to conduct an exhaustive investigation into the detailed variations of galaxy properties such as ellipticity, inclination, bulge fraction, etc. and their effect on the ability of SE to detect a given galaxy. What is important, however, is to note that the difference in detection rates of model galaxies based on a wide variety of real galaxies introduces a large scatter into the completeness estimates.  Inclusion of this scatter will be crucial in determining uncertainties in the completeness corrected number counts and will be discussed in the following section.

\subsection{Uncertainties}
\label{unc}

The total intensity of the 3.6 $\mu$m CIRB is computed by summing over the contributions from galaxies binned in 0.5 magnitude bins centered at each half magnitude:

\begin{equation}
I_{CIRB}(3.6 \mu m) =  \sum_{m} n(m) f(m)\, ,
\end{equation}
where n(m) is the number of galaxies per steradian at magnitude m and f(m) is the flux of the center of the magnitude bin.  A robust estimate of the error associated with estimating the CIRB by counting galaxies in IRAC images using Source Extractor and GIM2D requires an analysis of cosmic variance in addition to the Poisson shot noise in the counts, as well as the error in determining the completeness of the source detection and in determining the flux of each galaxy.  Uncertainties in the counts, completeness and flux are discussed in Sections~\ref{nofm} - Section~\ref{fofm}.  

\subsubsection{n(m)}
\label{nofm}

Uncertainties in n(m) are a combination of Poisson errors and cosmic variance.    
For $m \leq 16.0$, where n(m) is measured in our three regions, the measured fractional variance at each half magnitude: 

\begin{equation}
\frac{\delta n}{\langle n \rangle} = \frac{\sqrt{\langle (n - \langle n \rangle )^2 \rangle}}{\langle n \rangle}  \, , 
\end{equation}
where ${\langle n \rangle}$ is the average n(m) in the three regions, should include both the Poission noise and the cosmic variance since these fields are widely separated, allowing us to assume there is no spatial correlation between the counts in each field. These measured values, however, in some bins fall below the Poisson error based on the counts.  The values in Table 1 for $m \leq 16.0$, which are taken to be the uncertainties in those counts, are then the larger of the Poisson error or the measured fractional variance. 

In the deep fields (EGS, GOODSN), at fainter magnitudes, cosmic variance is comparable to the Poisson noise. While the fields are small in angular size, they are deep pencil beam surveys and thus sample a large volume of space. We can use a geometrical argument to estimate the cosmic variance in the deep fields.  The average of the SDSS photometric redshift estimates for galaxies with matches in the SDSS catalog is $z \approx 0.5$  These images probe much deeper than $z = 0.5$, however, we can use this to put an upper limit on the cosmic variance.  Taking the average depth $z \approx 0.5$ and a $\Lambda$CDM cosmology with $\Omega_{m} = 0.27$, $\Omega_{\Lambda}=0.73$, we reach a proper radial distance of $\approx 2700 h^{-1}$ Mpc ($h = $ H$_\circ$/100 km s$^{-1}$ Mpc$^{-1}$).  Breaking this up into pieces of length $8h^{-1}$ Mpc, this spans over three hundred $8h^{-1}$ Mpc slices. Taking $\sigma_{8} \approx 1$, where $\sigma_{8}$ is the relative density contrast in a spherical region of radius $8h^{-1}$ Mpc, and the variance proportional to $N_{fields}^{-\frac{1}{2}}$ where $N_{fields}$ is the number slices of length $8h^{-1}$ Mpc, this limits the fractional cosmic variance in the counts to less than about 6\%. This is consistent with the measured $\delta N/\langle N \rangle$ between the EGS and GOODS North fields of 5.1, 4.1 and 6.4\% at $m = 16.5, 17.0$ and $17.5$.  The error due to cosmic variance in the deep fields is then taken to be $\pm$ 5\%.  The total uncertainties in the number counts per magnitude at faint magnitudes, used in computing $\chi^2$ in the next section, are then, 
\begin{equation}
\sigma_n = \sqrt{\frac{n}{0.5\Omega}+(0.05n)^2} \, ,
\end{equation}
where n is the number per magnitude per square degree, $\Omega$ is the solid angle of the region in square degrees and the factor 0.5 comes from using half magnitude bins.  These are the values shown in Table~\ref{uncandcomp} for $m \geq 18$, and the quadrature sum of these uncertainties in the EGS and GOODS North is shown for 16.5 $\leq m \leq$ 17.5. 

\subsubsection{Completeness}
\label{dcomp}

As discussed above, completeness at each magnitude was determined by inserting approximately 40 different sample galaxy images into the science frames at 200 or 40 different random positions, depending on the image, and re-running the source detection and cross-correlating the new catalog with the old.  Thus, for each sample galaxy, $i$, we obtain a completeness estimate, $comp_i$, from the ratio of the number detected to the number inserted. The value of $comp_i$ varies widely with $i$ due to the different surface brightnesses and morphology of each sample galaxy.  We then run N$_{iter}$ = 4 or 10 iterations of this process, depending on whether we've inserted 200 or 40 copies of our sample galaxy. For the $j^{th}$ iteration, the completeness, $comp_j$, is taken to be the mean of the estimates for each galaxy and the completeness uncertainty, $\sigma_{comp,j}$, is conservatively taken to be the standard deviation of the distribution of the values $comp_i$.  The completeness at each half magnitude is then :
\begin{equation}
comp(m) = \langle comp_j \rangle \, , 
\end{equation} 
and
\begin{equation}
\sigma_{comp}(m) = \sqrt{\frac{1}{N_{iter}} \stackrel{N_{iter}}{\sum_{j=0}} \sigma_{comp,j}^2} \, . 
\end{equation} 
The completeness corrected, observed number counts them become:
\begin{equation}
n_{obs,corr}(m) = n_{obs}(m)/comp(m) \, ,
\end{equation}
with a total uncertainty of,
\begin{equation}
\sigma_{n,corr}^2(m) = \left[\left(\frac{\sigma_{n,obs}}{n_{obs}}\right)^2 + \left(\frac{\sigma_{comp}}{comp}\right)^2\right]\left(\frac{n_{obs}}{comp}\right)^2 
\end{equation} 

\subsubsection{f(m)}
\label{fofm}

\begin{figure}
\epsscale{.80}
\plotone{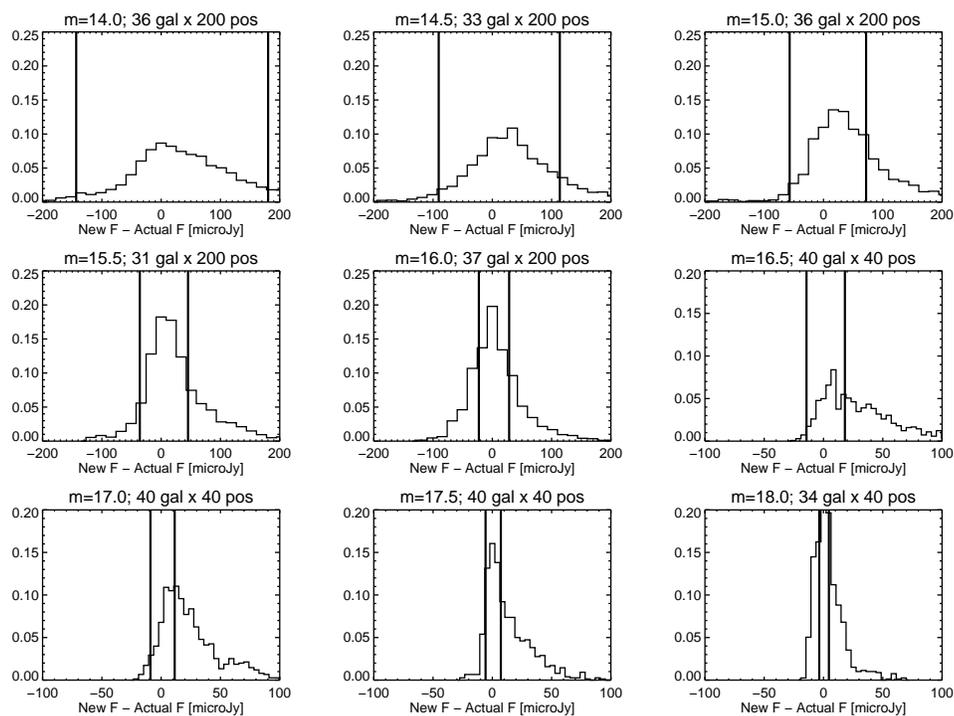}
\caption{Results of inserting and re-fitting the indicated number of galaxy images.  New F is the re-fit flux and Actual F is the total of the pixel values actually inserted.  Indicated are the number of sample galaxies and the number of random positions at which they were inserted.  Vertical lines represent the edges of the 0.5 magnitude bins.} 
\label{hist}
\end{figure}

Once the sample galaxies were inserted into the science frame and re-detected for the incompleteness estimates, for the last run, the entire analysis pipeline was re-run on those sample galaxies that were successfully detected as new sources within 1.5 pixels of the insertion coordinates.  After the sample galaxies were re-fit, the new fluxes determined by GIM2D were compared with the actual flux of the inserted galaxy.  Histograms of the total flux from the re-fit, with the total of the inserted pixel values subtracted, are shown for each half magnitude from $m = 14$ to $m = 18$ in Figure ~\ref{hist}. The histograms are normalized to a total of one.  Vertical lines in the plots show the widths of the 0.5 magnitude bins.  One can see from the histograms that the modes of the distributions are near zero, i.e. the most likely value of the re-fit flux is the actual flux.  However, we also see that many of the new fluxes spill out of the original bins into neighboring magnitude bins with a clear skew toward overestimation of the flux.  Given this information, however, we use a Monte Carlo Markov Chain (MCMC) simulation of our analysis to put constraints on the possible real distributions of n(m) that can give rise to counts we observe.  

The model for the counts of galaxies as a function of galaxy flux, n(f) [mag$^{-1}$ deg$^{-2}$], is a simple broken power law\footnote{In principle, one can always generate a very large CIRB by adding terms to $n(f)$ that give large counts for unobservably small fluxes.  Our simple form for $n(f)$ rules out this possibility a priori.}:
\begin{equation}
\label{model_func}
n(f) = \frac{A f^{-\alpha}}{(1+(f_\circ/f)^\beta)}  \, . 
\end{equation}
where A, $\alpha$, $\beta$ and $f_\circ$ are adjustable parameters and the bright end counts are Euclidian if $\alpha = 1.5$.  Two prior assumptions were made.  The first is that the integrated flux converges, achieved by requiring a faint end slope $\alpha - \beta < 1$.  The second is a prior on the bright end slope, $\alpha$, such that before subjecting the model to the data, we assume $\alpha = 1.5 \pm 0.25$.  The likelihood function is given by $L = e^{-0.5\chi^2}$ where  
\begin{equation}
\chi^2 = \sum_{m} (n_{model_{obs}}(m) - n_{obs,corr}(m))^2/\sigma_{n,corr}^2 \, .
\end{equation}
The posterior probability is then given by:
\begin{equation}
\label{posterior}
P(A,\alpha,\beta,f_\circ|n_{obs,corr}(m)) = e^{-0.5\chi^2} p(\alpha - \beta) p(\alpha) \, , 
\end{equation}
where
\begin{equation}
p(\alpha - \beta) = \cases{1, &for $\alpha - \beta < 1$ ,\cr 0, &for $\alpha - \beta > 1$ ,\cr} 
\end{equation}
and
\begin{equation}
p(\alpha) = \mbox{exp}\left(-0.5\left(\frac{\alpha - 1.5}{0.25}\right)^2\right) \, .
\end{equation}
The parameter space is then explored using Markov Chains (see \citet{sm91} or \citet{bind97}), starting with a various sets of initial parameters. The MCMC simulation proceeds as follows:
\begin{enumerate}
\item Compute the modeled ``real'' counts at each half magnitude for $10.5 < m < 19.5$, given the current set of parameters using Equation~\ref{model_func}.
\item Compute the modeled ``observed'' counts by multiplying the modeled ``real'' counts vector $\bf n\rm_{model_{real}}$ by the transfer matrix, \bf h\rm, and summing the resulting counts at each flux, $j$, i.e. $n_{model_{obs},j} = \sum_{i} n_{model_{obs},i} h_{ij}$.  The $i^{th}$ row of the matrix, one for each half magnitude bin, is given by the histograms in Figure~\ref{hist} with the actual flux at the bin center added back in. Histograms for $m < 14$ and $m > 18$ are the $m = 14$ and $m = 18$ histograms respectively, scaled by the central flux of the appropriate bin. 
\item The resulting modeled ``observed'' counts are spread over the full range of fluxes, $j$, and must then be binned into the same 19 half magnitude bins as the observed counts. We can then compute the posterior probability as given by Equation~\ref{posterior}.
\item Generate a random change in the parameters $\delta$A, $\delta\alpha$, $\delta\beta$ and $\delta f_\circ$.
\item Repeat Steps 1 - 3
\item If the probability of the new parameters, $P_{new}$, is higher than that of the old parameters, $P_{old}$, the new parameters are accepted and the process is repeated from Step 4
\item If $P_{new} < P_{old}$, the new parameters are accepted with the probability $P_{new}/P_{old}$ and the process is repeated from Step 4.  This facilitates exploration of the entire parameter space by sometimes accepting a downhill step. Otherwise, the new parameters are rejected and the process is repeated from Step 4.
\end{enumerate}
Several chains were run with 200,000 steps each.  

\section{Results}
\label{results}

\begin{figure}
\epsscale{.80}
\plotone{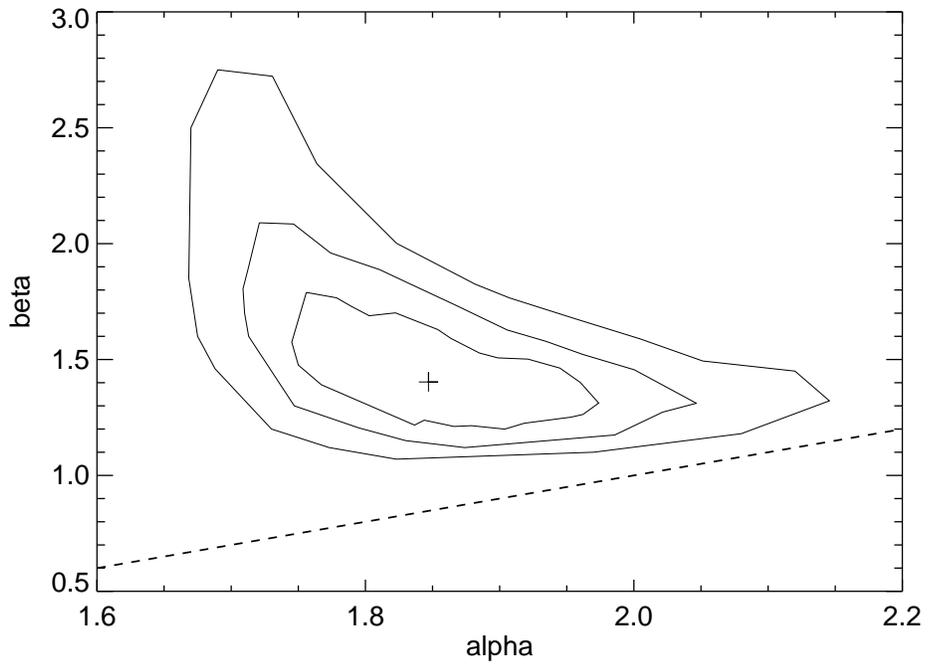}
\caption{68\%, 90\% and 99\% confidence limits in the alpha-beta plane.  The dotted line shows the convergence prior ($\alpha - \beta < 1$) and the cross indicates most probable values.  } 
\label{alphabeta}
\end{figure}

\begin{figure}
\epsscale{.80}
\plotone{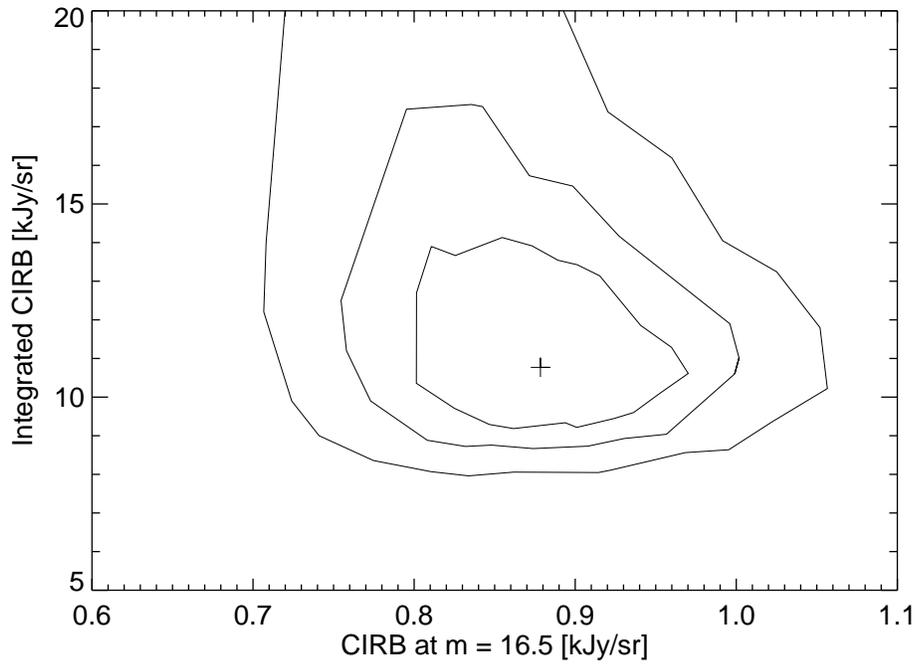}
\caption{68\%, 90\% and 99\% confidence limits in the CIRB(m=16.5)-full CIRB plane. The CIRB at m = 16.5 is clearly constrained, while the full CIRB is not.  The cross indicates the most probable values.} 
\label{c165}
\end{figure}

\begin{figure}
\epsscale{.80}
\plotone{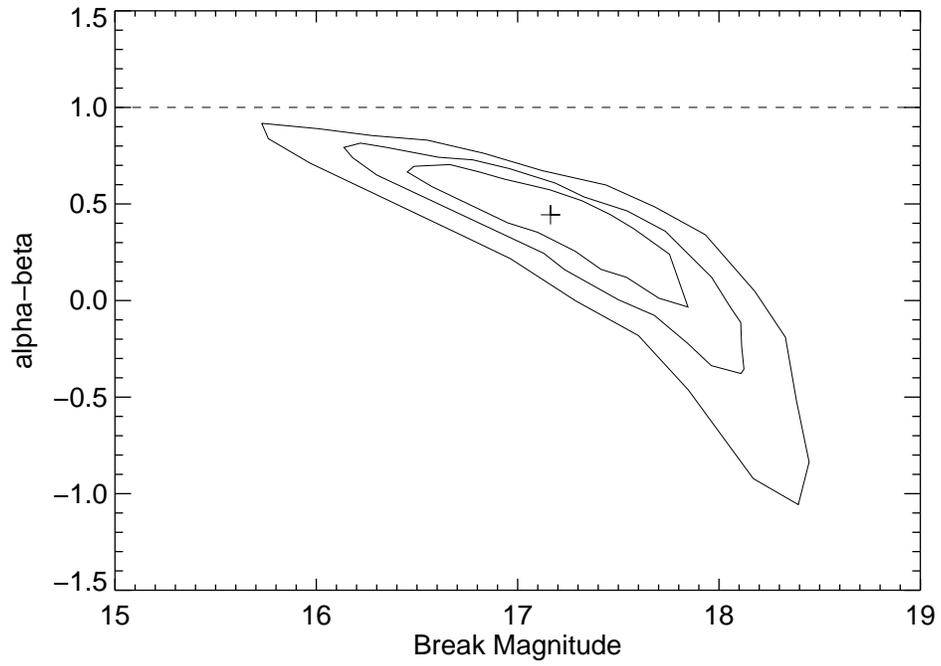}
\caption{68\%, 90\% and 99\% confidence limits on the faint end slope in the power law model of the galaxy counts vs. the magnitude at which the break occurs. The cross indicates the most probable values.} 
\label{fo}
\end{figure}

\begin{figure}
\epsscale{.80}
\plotone{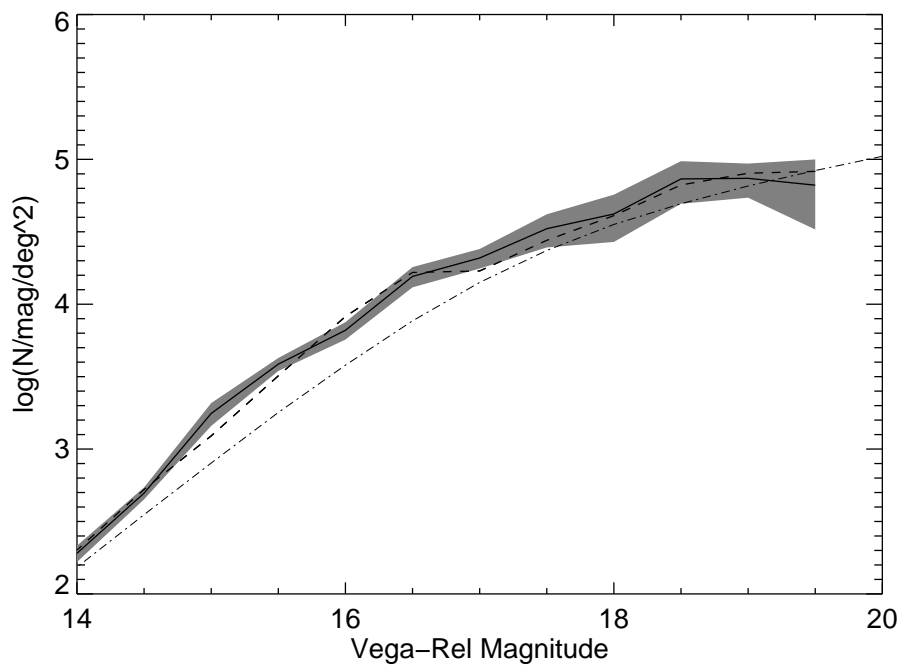}
\caption{Using the averaged counts from our three regions down to m = 16, averaged counts from the Groth Strip and GOODS North from 16.5 $\leq$ m $\leq$ 17.5 and deep GOODS North counts for m $\geq$ 18, we obtain completeness corrected observed profile-fit galaxy counts at each magnitude at 3.5 $\mu$m (solid).  The shaded area indicates 1$\sigma$ error bars on the corrected counts.  The dashed curve shows the modeled ``observed" counts. Dot-dashed curve shows the most probable model for the ``real'' counts.  } 
\label{modeln}
\end{figure}

\begin{figure}
\epsscale{.80}
\plotone{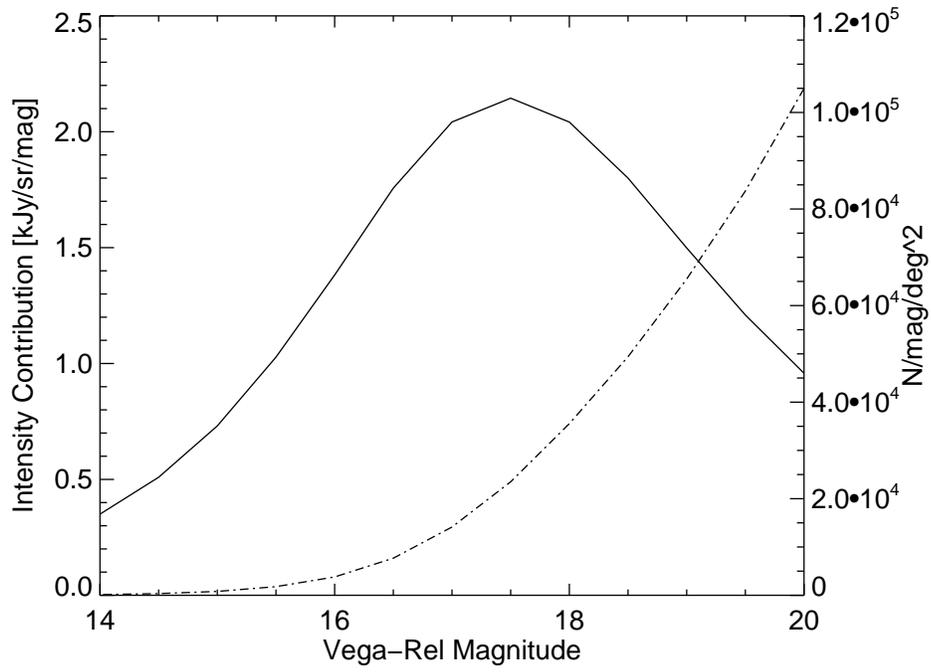}
\caption{Using the maximum likelihood model for the ``real'' galaxy counts (dot-dashed), we obtain a total intensity contribution from galaxies as a function of magnitude at 3.5 $\mu$m (solid).} 
\label{modeln_cirb}
\end{figure}

Figure~\ref{likelihood} shows the posterior probability distribution and the resulting CIRB for the models with the highest probability in each bin.  
The distribution gives a most probable contribution from galaxies to the CIRB of $10.8^{+2.1}_{-1.1}$ kJy sr$^{-1}$ ($9.0^{+1.7}_{-0.9}$ nW m$^{-2}$ sr$^{-1}$) where the asymmetric range is given by the full width of the histogram where the probability falls from its maximum by a factor $e^{-\frac{1}{2}}$. Parameters for the most probable model are given in Table~\ref{params}. One can see that even given the prior on $\alpha$, the most probable bright end slope for the counts is steeper at $\alpha = 1.85$.  Figure~\ref{alphabeta} shows the 68\%, 90\% and 99\% confidence limits on the parameters $\alpha$ and $\beta$. The dashed line shows the convergence prior.  Figure~\ref{c165} shows the same confidence limits on the CIRB intensity at $m = 16.5$ and the integrated CIRB. We can see that the intensity of galaxies around $m = 16.5$ is well constrained, while the integrated CIRB intensity is not on the bright end.  Figure~\ref{fo} shows the same confidence limits on the faint end slope in the power law model of the counts ($\alpha - \beta$) and the magnitude at which the break occurs (Break Magnitude = $-2.5\log(f_\circ/F_\circ)$) where $F_\circ = 277.5$ Jy is the flux density of Vega. Indicated by the cross in these three figures are the most probable values for those parameters. Counts and intensities as a function of magnitude derived from the most probable set of parameters are shown in Figures~\ref{modeln} and \ref{modeln_cirb}.  Figure~\ref{modeln} shows the completeness corrected observed counts, the model ``observed'' counts and the model ``real'' counts, versus magnitude.  The CIRB intensities resulting from the raw galaxy counts, the completeness corrected counts and the most probable model are shown in Table~\ref{corr}.  Figure~\ref{modeln_cirb} shows the modeled ``real'' counts and the resulting contribution to the CIRB at each magnitude.  

\subsection{Aperture Photometry}

\begin{figure}
\epsscale{.80}
\plotone{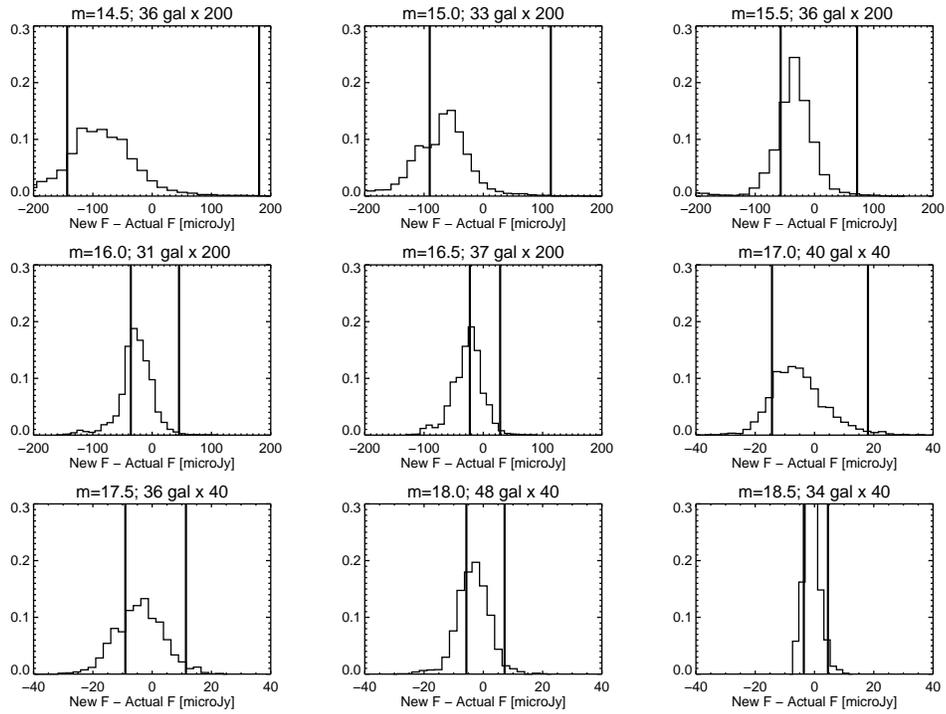}
\caption{Histograms, arranged as in Figure~\ref{hist}, but here, New F are SExtractor aperture fluxes and the magnitude at which the histograms are centered are shifted one half-magnitude fainter since the galaxies used were selected by their \it profile-fit \rm magnitudes.} 
\label{hist_aper}
\end{figure}

\begin{figure}
\epsscale{.80}
\plotone{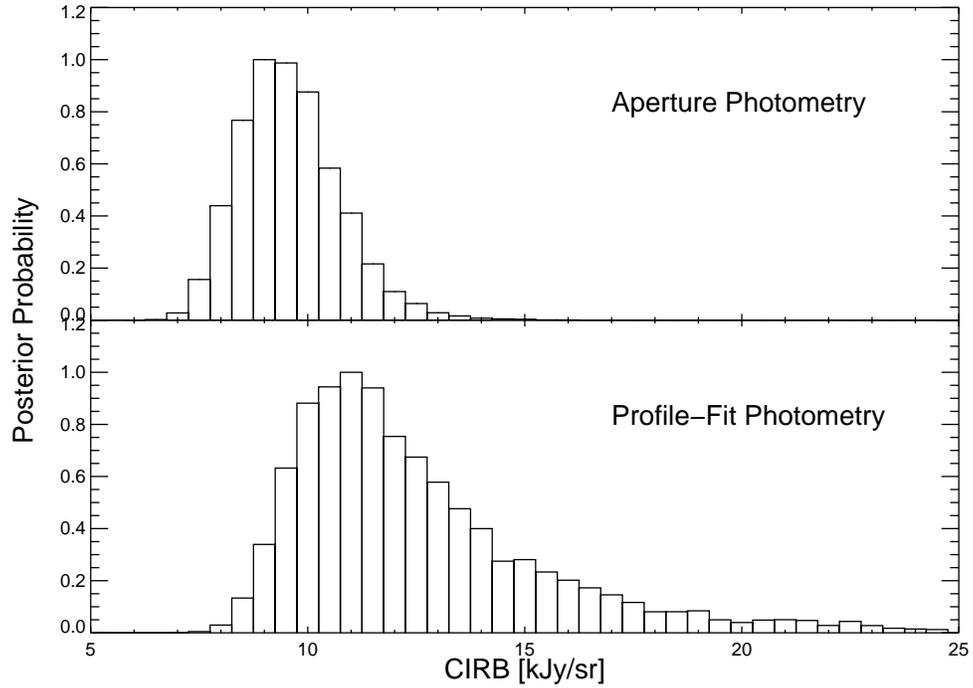}
\caption{Scattering the model counts, now with the aperture photometry histograms in Figure~\ref{hist_aper}, the MCMC analysis results in a CIRB of $9.2^{+1.2}_{-0.7}$ kJy sr$^{-1}$  ($7.6^{+1.0}_{-0.6}$ nW m$^{-2}$ sr$^{-1}$), where the asymmetric range is given by the full width where the probability drops from its maximum by a factor $e^{-\frac{1}{2}}$. The profile-fit results are shown in the background for comparison.} 
\label{likelihood_aper}
\end{figure}

\begin{figure}
\epsscale{.80}
\plotone{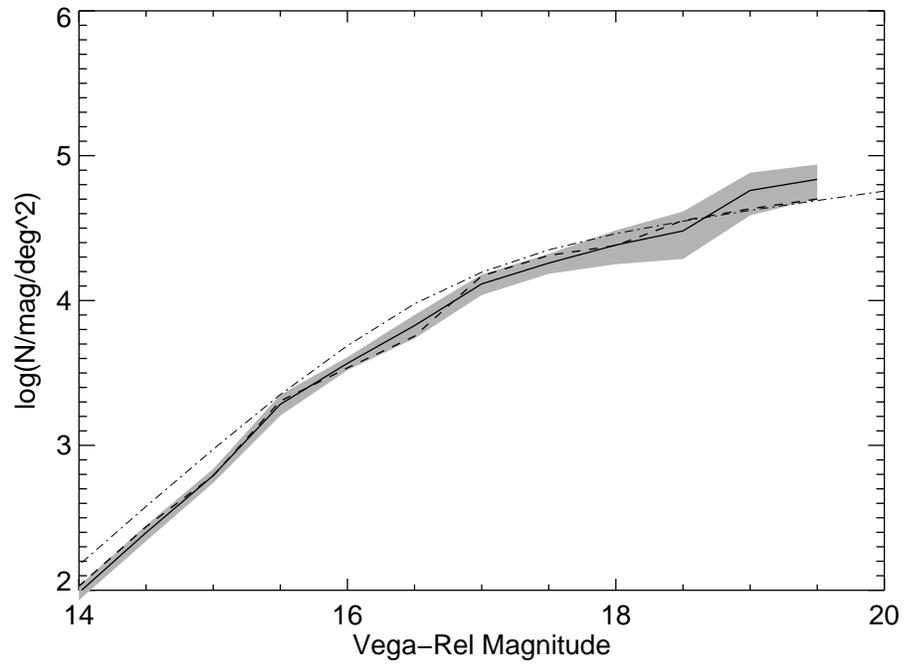}
\caption{Curves are the same as in Figure~\ref{modeln}, but here all counts are determined using aperture photometry.} 
\label{modeln_aper}
\end{figure}

\begin{figure}
\epsscale{.80}
\plotone{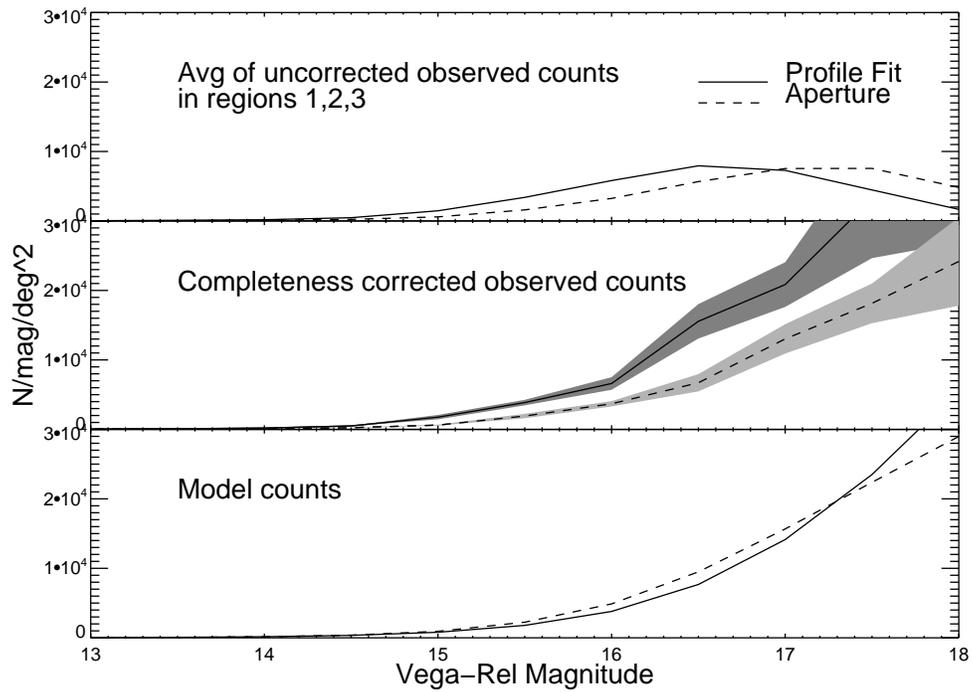}
\caption{Profile-Fit vs. Aperture photometry. Top Panel: Average of the uncorrected observed counts in regions 1, 2 \& 3 as seen in Figure~\ref{aper}.  Middle Panel: Completeness corrected observed counts as seen in Figures~\ref{modeln} \& \ref{modeln_aper}.  Bottom Panel: Modeled ``real'' counts as determined by the MCMC simulation using both profile-fit and aperture photometry.} 
\label{bothcounts}
\end{figure}

As a check, we ran the same MCMC analysis using counts obtained with SE aperture photometry.  Aperture fluxes of the galaxies inserted at each magnitude for the completeness testing were compared to the total of the pixel values actually inserted for each source and binned as before.  The resulting comparison is shown in Figure~\ref{hist_aper} where the magnitude at which the histogram is centered is shifted by one half-magnitude with respect to those in Figure~\ref{hist}.  This shift, along with a similar shift in the completeness estimates, is due to the use (in measuring the completeness and flux uncertainties) of galaxies selected by their \it profile-fit \rm magnitudes.  We have seen that these are, on average, 0.5 magnitude brighter than the SE aperture magnitudes.  We can see from Figure~\ref{hist_aper} that aperture photometry tends to underestimate the flux.  The MCMC analysis was run again with these histograms used in place of those in Figure~\ref{hist} to scatter the modeled ``real'' counts.  The resulting probability distribution for the intensity of the CIRB is shown in Figure~\ref{likelihood_aper}.  Here we have a somewhat tighter distribution, but we still have a long non-Gaussian tail extending to larger intensities. The full width of the histogram at $P = e^{-\frac{1}{2}} P_{max}$ gives a CIRB of $9.2^{+1.2}_{-0.7}$ kJy sr$^{-1}$ ($7.6^{+1.0}_{-0.6}$ nW m$^{-2}$ sr$^{-1}$) and the best fit parameters are shown in the second column of Table~\ref{params}.  The resulting model ``observed'' counts, along with the model ``real'' counts and the completeness corrected observed aperture photometry counts are shown in Figure~\ref{modeln_aper}. The CIRB intensities resulting from the raw aperture photometry galaxy counts, the completeness corrected aperture counts and the resulting most probable model are shown in Table~\ref{corr}.  The remaining 1$\sigma$ difference between the aperture and profile-fit intensities is approximately what one would expect, given that the two measurements are not independent, and is due mainly to the difference in the faint end slope of the model counts.  This slope is determined by the most uncertain faint counts as seen in the middle panel of Figure~\ref{bothcounts}.  However, as shown in the bottom panel of Figure~\ref{bothcounts}, the MCMC analysis does give a more consistent picture of the counts determined by the two methods.  We report here the profile-fit results, over those from aperture photometry, based on the ratio of the maximum probabilities in each case: $P_{max}(aper)/P_{max}(profile) \approx 0.04$.

\section{Discussion}

\citet{elw01b} extrapolated the SDSS determined optical luminosity function into the NIR and found it to be a factor of 2.3 higher than that determined with 2MASS by \citet{koch01} and \citet{cole01}. Similar discrepancies of factors of 1.5 and 2.1 were also seen among optical luminosity functions when comparing the SDSS luminosity function with those determined using the 2 degree Field Galaxy Redshift Survey \citep{folk99} and the Las Campanas Redshift Survey \citep{lin96}.  \citet{bla01} showed that the optical discrepancy was explained by the use of larger apertures by the SDSS.  The NIR discrepancy was not explained, but \citet{elw01b} suggested that inclusion of the faint fringes of galaxies using larger apertures in the NIR could be at least a partial explanation for the discrepancy seen between the SDSS extrapolated and 2MASS luminosity functions.  The \citet{elw01} DIRBE minus 2MASS subtraction, recently extended to 40 new regions of the sky in \citet{me07a} resulted in a detection of a statistically significant isotropic background of 15.6 $\pm$ 3.3 kJy sr$^{-1}$ (13.3 $\pm$ 2.8 nW m$^{-2}$ sr$^{-1}$) at 3.5 $\mu$m.  \citet{faz04}, using custom aperture photometry, determined a total contribution of resolved galaxies to the CIRB of only 6.5 kJy sr$^{-1}$ (5.4 nW m$^{-2}$ sr$^{-1}$), less than half of that value.  Hence, we set out to include the the faint fuzzy fringes of the galaxies.  Aperture photometry is limited, especially in crowded fields where the edges of the galaxies overlap and large apertures can not be used.  Here we have employed profile-fit photometry and raised the \citet{faz04} hard lower limit and replaced it with a Gaussian lower limit using counts, obtained with GIM2D, of galaxies in 3 DIRBE Dark Spots, as well as deep images of the Extended Groth Strip and the GOODS North field.   We are able, using the machinery of MCMC simulation, to obtain a model for the distribution of galaxies, $n(m)$, that gives a most probable total galactic contribution to the CIRB of $10.8^{+2.1}_{-1.1}$ kJy sr$^{-1}$ ($9.0^{+1.7}_{-0.9}$ nW m$^{-2}$ sr$^{-1}$). These values are consistent, for the first time, with previous determinations of the CIRB in the L band shown in Table~\ref{previous}, excepting the \citet{faz04} lower limit, which, based on the 1$\sigma$ lower bound from the width of the probability histogram at $P = P_{max} e^{-0.5}$, is 3.5$\sigma$ below the most probable intensity determined here.

\begin{figure}
\epsscale{.80}
\plotone{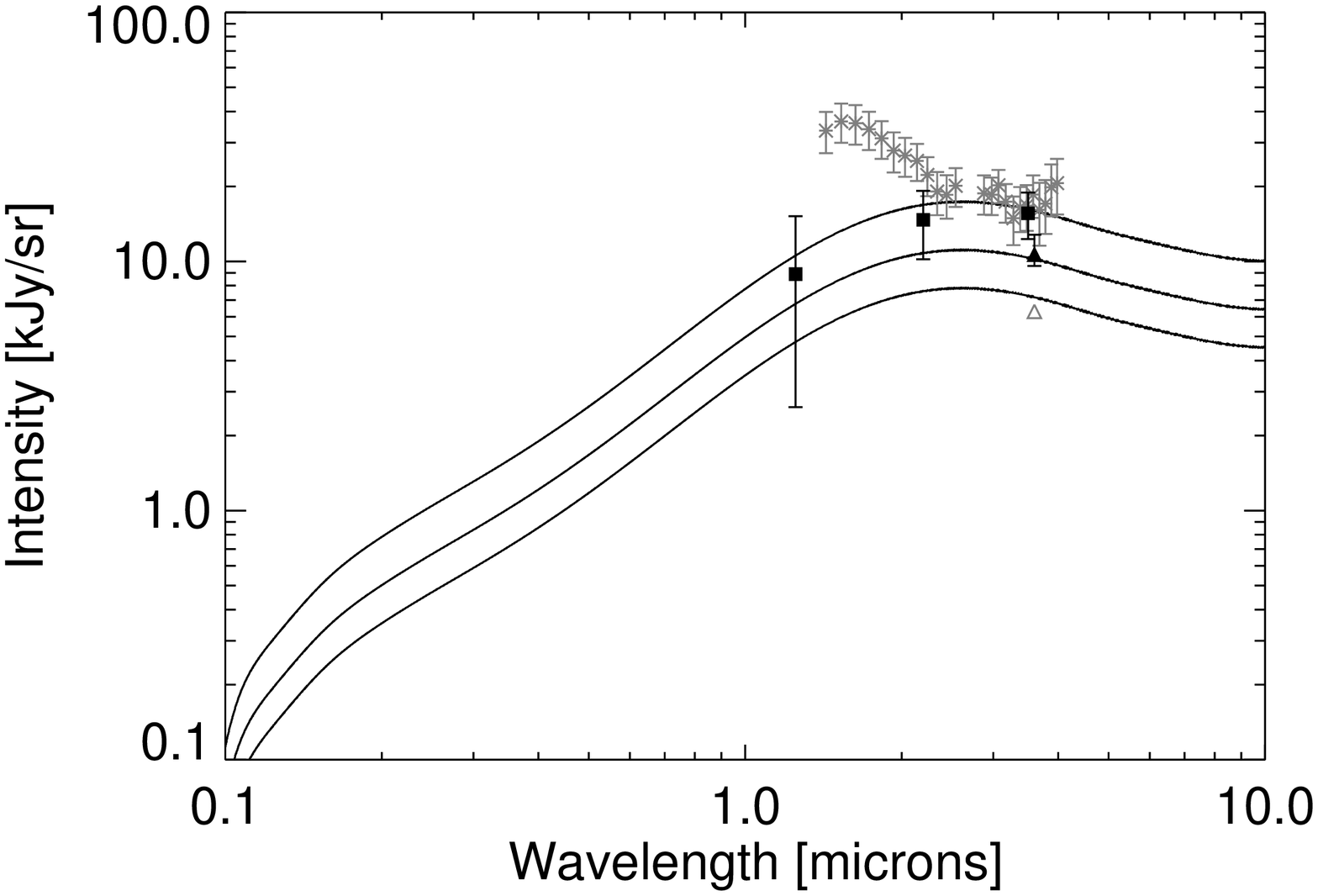}
\caption{The filled triangle is the $10.8^{+2.1}_{-1.1}$ kJy sr$^{-1}$ ($9.0^{+1.7}_{-0.9}$ nW m$^{-2}$ sr$^{-1}$) value reported here. Squares are the CIRB values reported in \citet{me07a}.  Gray stars are \citet{mat05} values from IRTS observations.  The open gray triangle is the \citet{faz04} lower limit at 3.5$\mu$m.  For comparison, the upper and lower black curves are the P1.0 and P0.45 models used by \citet{aha06} to estimate the attenuation of TeV $\gamma$-rays by the CIRB. P1.0 was normalized by \citet{aha06} to fit the 1.25-3.5 $\mu$m values from \citet{dwe98} and \citet{elw01}.  P0.45 is the P1.0 scaled down by a factor of 0.45, which was required, using this shape for the CIRB, to give blazar spectra with power law spectral indices of at least 1.5.. Also shown is P0.64 which fits the value reported here.} 
\label{cirb}
\end{figure}

A third constraint on the CIRB comes from the attenuation of TeV $\gamma$-rays by $e^+e^-$ pair production through collisions with CIRB photons. \citet{dwe05} derived the attenuating effect on $\gamma$-ray photons from the Blazar PKS 2155-304 of the 1$\mu$m spike in the CIRB suggested by the \citet{mat05} IRTS measurements shown in Figure~\ref{cirb}. That analysis showed that such a spike is unlikely to be of extragalactic origin and that the CIRB can not be significantly higher than the intensity derived from galaxy counts.  H.E.S.S. observations of the blazars H2356-309 and 1ES 1101-232 by \citet{aha06} also suggest that the CIRB can not be much higher than previously determined lower limits from galaxy counts.  The \citet{aha06} P1.0 Extragalactic Background Light (EBL) model, used for estimating $\gamma$-ray attenuation gives 11, 18  and 16 kJy sr$^{-1}$ at 1.25, 2.2 and 3.5 $\mu$m (26, 25 and 14 nW m$^{-2}$ sr$^{-1}$) and is shown as the top curve in Figure~\ref{cirb}.  Assuming a power law blazar spectrum ($dN/dE$ [photons cm$^{-2}$ s$^{-1}$ TeV$^{-1}$] $\propto$ $E^{-\Gamma}$) this model results in a power law index for the spectrum of the blazar 1ES 1101-232 of $\Gamma$ = -0.1. The model must be scaled down by a factor of 0.45 to the P0.45 model (bottom curve Figure~\ref{cirb}) to give a power law index of at least 1.5, which is considered by \citet{aha06} to be the lowest acceptable value.  The middle curve in the Figure is P0.64 which is the P1.0 model scaled to fit the  $10.8^{+2.1}_{-1.1}$ kJy sr$^{-1}$ ($9.0^{+1.7}_{-0.9}$ nW m$^{-2}$ sr$^{-1}$) value reported here.  \citet{map06} show however that an EBL model, also based on the \citet{elw01} values at 1.25 and 2.2 $\mu$m, but with a steeper decline from 4 to 10 $\mu$m results in a power law index of $\Gamma$ = +0.5.  They consider $\Gamma$ = 0.6 to be the lowest acceptable index based on physical considerations and suggest then that while the DIRBE minus 2MASS CIRB values at 1.25 and 2.2 $\mu$m from \citet{elw01} and hence the 3.5 $\mu$m values reported here and elsewhere in Table~\ref{previous} do require a hard spectrum, and the lower limits from galaxy counts are favored, they can not be ruled out based on the current H.E.S.S. data.  

\citet{vas00} showed that it is difficult to untangle attenuation due to the EBL from unknown intrinsic properties of blazars, including the total blazar flux or an intrinsic deviation from a pure power law spectrum.   He showed also that the increased sensitivity of H.E.S.S., VERITAS and MAGIC, leading to detections of more blazars at more redshifts now and in the near future, and the expanded energy range of these instruments compared with their predecessors are crucial to efforts to decouple those effects and extract information about both the EBL and TeV blazars.  The MAGIC collaboration \citep{tesh07} claims a detection of 3C 279 at a redshift of 0.5.  If we are able to detect TeV sources at such large distances, we may be limited to very few CIRB photons as the interaction between $\gamma$-ray and CIRB photons can put non-trivial limits on the spectrum of the EBL. However, given the convergence toward the higher CIRB intensities measured through galaxy counting with the faint fringes of the galaxies included and the direct measurements in Table~\ref{oldvals}, the intensity of the CIRB at 3.6$\mu$m may be seen as an important anchor for EBL models which will allow exploitation of this interaction in determining intrinsic blazar properties and perhaps probing certain aspects of fundamental physics such as a boost to the propagation of TeV photons by axion production.  While the implications of $\gamma$-ray attenuation by CIRB photons are still limited by the small number of observed sources, this indirect, independent probe of the CIRB will improve as more blazars are observed by H.E.S.S., VERITAS and MAGIC.

\begin{figure}
\epsscale{.80}
\plotone{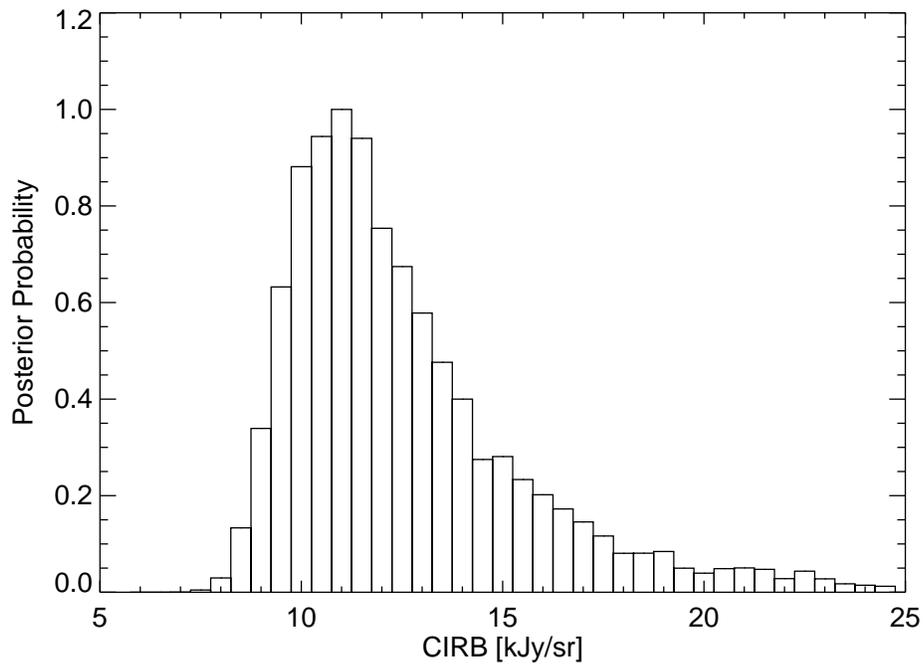}
\caption{Results of the MCMC simulation; the distribution of relative probabilities for models resulting in a given CIRB.  The most probable model from profile-fitting results in a CIRB of $10.8^{+2.1}_{-1.1}$ kJy sr$^{-1}$ ($9.0^{+1.7}_{-0.9}$ nW m$^{-2}$ sr$^{-1}$), where the asymmetric range is given by the full width where the probability drops from its maximum by a factor $e^{-\frac{1}{2}}$.} 
\label{likelihood}
\end{figure}

Taking this new value from profile-fitting of $10.8^{+2.1}_{-1.1}$ kJy sr$^{-1}$ ($9.0^{+1.7}_{-0.9}$ nW m$^{-2}$ sr$^{-1}$) to be the fiducial value for the contribution of galaxies to the CIRB, we are left with only a 1.2$\sigma$ difference between the values determined via galaxy counts and those determined via the direct detection method of ``DIRBE minus 2MASS'' and others in determining the L-band CIRB. Even the more conservative aperture counts are now only 1.8$\sigma$ lower than the direct measurements.  These numbers are significantly less that the 2.8$\sigma$ discrepancy between direct measurements and the \citet{faz04} lower limit of 6.5 kJy sr$^{-1}$ (5.4 nW m$^{-2}$ sr$^{-1}$). It is also important to note, however, that the results from profile-fitting do not strongly constrain the intensity of the CIRB on the bright end.  The probability distribution in Figure~\ref{likelihood} is non-Gaussian and the width where the probability drops from its maximum by a factor $e^{-2}$ already allows a CIRB of 17.1 kJy sr$^{-1}$ (14.2 nW m$^{-2}$ sr$^{-1}$). Galaxy counts using both aperture photometry and profile-fit photometry still bound the CIRB from below, but even with a restricted broken power law model for the counts, we are able to find solutions where the CIRB goes to large intensities and it is unnecessary to invoke an additional component with large numbers at unobservably faint fluxes. Gamma-ray attenuation may still favor a lower intensity, but we do seem to be moving toward a consistent picture of the 3.5-3.6 $\mu$m CIRB from direct detection and galaxy counting. The goal then will be to continue to drive down the uncertainties in both the galaxy counting and direct measurements.  More data from \it Spitzer \rm can help here.  One of the limitations of this study is that we have wide shallow surveys, but the deep images only survey a tiny solid angle.  Multiple deep wide surveys would allow more robust galaxy counts at the important magnitudes $16 < m < 20$.  At 3.6 $\mu$m these can be done during the post-cryogen extended \it Spitzer \rm mission when competition with other instruments for telescope time will be eliminated.  This method is limited, however, by confusion not only due to the IRAC pixel size, but more fundamentally due to the real confusion between the faint fringes of galaxies in the deep sky.   The uncertainties inherent in all of these methods means it will likely take an improvement of the zodiacal light models to make to make a significant improvement on the precision of these measurements.   The 4 kJy sr$^{-1}$ (3.4 nW m$^{-2}$ sr$^{-1}$)difference between the \citet{wr98} and \citet{kel98} models indicates the level of uncertainty in the zodiacal light subtraction.  A space mission to directly observe the zodiacal light, of the type suggested in \citet{me07a} may be the only way reduce the dominant source of uncertainty in measuring the intensity of the Cosmic InfraRed Background.

\acknowledgments

This work is based on observations made with the Spitzer Space Telescope, which is operated by the Jet Propulsion Laboratory, California Institute of Technology under contract with NASA.  Support for this work was provided by NASA through an award issued by JPL/Caltech.

\clearpage
\begin{deluxetable}{ccccccc}
\tablecaption{DIRBE Dark Spot Coordinates \label{coords}}
\tablewidth{0pt}
\tablehead{
\colhead{Region} & \colhead{l} & \colhead{b} & \colhead{$\lambda$} & \colhead{$\beta$} & \colhead{RA} &\colhead{DEC} }
\startdata
      1  &     92.9  &     +70.2  &      181.9  &      +49.5 &      13h45m33.6s &      +43d35m59.8s\\
      2  &     318.4  &     +76.9  &      189.3  &      +19.7 &      13h05m49.7s &      +14d25m11.9s\\
      3  &     108.9  &     +46.4  &      153.0  &      +70.7 &      14h36m23.2s &      +67d46m18.0s\\
\enddata

\tablecomments{Galactic and ecliptic (J2000) coordinates are given in degrees.  Equatorial coordinates are J2000. }
\end{deluxetable}

\clearpage

\clearpage
\begin{deluxetable}{ccc}
\tabletypesize{\scriptsize}
\tablecaption{Uncertainties and Completeness\label{uncandcomp}}
\tablewidth{0pt}
\tablehead{
\colhead{Central Mag\tablenotemark{a}} & \colhead{$\sigma_n$} & \colhead{Completeness} }
\startdata
      10.5  &     $\pm$ 141\%  & 100 $\pm$ 5\%  \\
      11.0  &     $\pm$ 59\%  & 100 $\pm$ 5\%  \\
      11.5  &     $\pm$ 59\%  & 100 $\pm$ 5\%  \\
      12.0  &     $\pm$ 48\%  & 100 $\pm$ 5\%  \\
      12.5  &     $\pm$ 30\%  & 100 $\pm$ 5\%  \\
      13.0  &     $\pm$ 18\%  & 100 $\pm$ 5\%  \\
      13.5  &     $\pm$ 10\%  & 100 $\pm$ 5\%  \\
     14.0  &     $\pm$ 7.5\%  & 93 $\pm$ 10\%  \\
      14.5  &     $\pm$ 9.0\%  & 95 $\pm$ 3\%  \\
      15.0  &     $\pm$ 11.0\%  & 83 $\pm$ 12\%  \\
      15.5  &     $\pm$ 8.9\%  & 88 $\pm$ 5\%   \\
      16.0  &     $\pm$ 4.0\% &   88 $\pm$ 12\%   \\
      16.5  &     $\pm$ 9.5\% &    74 $\pm$ 10\%   \\
      17.0  &     $\pm$ 8.3\%  &   73 $\pm$ 10\%  \\
      17.5  &     $\pm$ 6.3\% &    68 $\pm$ 17\%  \\
      18.0  &    $\pm$ 7.0\%  &    71 $\pm$ 25\%  \\
      18.5  &    $\pm$ 6.6\%  &    47 $\pm$ 15\%  \\
      19.0  &    $\pm$ 6.9\%  &    43 $\pm$ 11\%  \\
      19.5  &    $\pm$ 7.4\%  &    37 $\pm$ 18\%  \\
\enddata

\tablenotetext{a}{Magnitudes shown here are profile-fit magnitudes.}
\end{deluxetable}

\clearpage
\begin{deluxetable}{ccc}
\tablecaption{Most Probable Parameters\label{params}}
\tablewidth{0pt}
\tablehead{
\colhead{} &\colhead{Profile-Fit} & \colhead{Aperture}  }
\startdata
A  			&     $2.3 \times 10^{-4}$   & $0.5 \times 10^{-4}$\\
$\alpha$		&     $1.85$   		& $2.06$\\
$\beta$ 		&     $1.40$  		& $1.75$\\
$m(f_\circ)$  		&     17.2  	& 16.6 \\
\enddata
\end{deluxetable}

\clearpage
\begin{deluxetable}{lccc}
\tablecaption{CIRB Corrections [kJy sr$^{-1}$]\tablenotemark{a} \label{corr}}
\tablewidth{0pt}
\tablehead{
\colhead{Photometry} & \colhead{$\sum n(m)f(m)$} &\colhead{$\sum (n(m)/comp(m))f(m)$} & \colhead{$\sum n_{model}(m)f(m)$}  }
\startdata
Profile Fit &  	9.3		&     12.9   & $10.8^{+2.1}_{-1.1}$\\
Aperture &  	5.3		&     7.1   & $9.2^{+1.2}_{-0.7}$\\
\enddata

\tablecomments{The non-Gaussian nature of the flux errors, seen in Figures~\ref{hist} and \ref{hist_aper}, makes a determination of the errors in the first two numerical columns difficult, making necessary the MCMC simulation that provides the above 1$\sigma$ confidence limits on the modeled CIRB.}
\tablenotetext{a}{For comparison, $\nu I_\nu$ [nW m$^{-2}$ sr$^{-1}$] = $\frac{3}{\lambda [\mu m]}I_\nu$ [kJy sr$^{-1}$]}
\end{deluxetable}

\clearpage 

\begin{deluxetable}{lcccl}
\tablecaption{Previous determinations of the L-band\tablenotemark{a} CIRB [kJy sr$^{-1}$]\tablenotemark{b} \label{oldvals}\label{previous}}
\tablewidth{0pt}
\tablehead{
\colhead{Authors} & \colhead{CIRB} & \colhead{Zodi-Model\tablenotemark{c}} }
\startdata
\citet{dwe98}	& 11.6 $\pm$ 3.4& \citet{kel98}\\
\citet{gor00}	& 12.8 $\pm$ 3.8& \citet{wr98}\\
\citet{wrr00}	& 14.4 $\pm $3.7 & \citet{wr98}\\
\citet{wrj01}	& 16.1 $\pm$ 4 & \citet{wr98}\\
\citet{mat05}	& 16.9 $\pm $ 3.5 & \citet{kel98}	\\
\citet{me07a} 	& 15.6 $\pm$ 3.3\ & \citet{wr98}\\
\citet{faz04}	& $>$ 6.5 & N.A. (galaxy counts)\\
This Work & $10.8^{+2.1}_{-1.1}$ & N.A. (galaxy counts)\\
\enddata

\tablenotetext{a}{Foreground subtraction intensities were measured in the DIRBE 3.5 $\mu$m band while galaxy count intensities were measured in the \it Spitzer \rm 3.6 $\mu$m band.  These results do not distinguish between these nearly identical bands.}
\tablenotetext{b}{For comparison, $\nu I_\nu$ [nW m$^{-2}$ sr$^{-1}$] = $\frac{3}{\lambda [\mu m]}I_\nu$ [kJy sr$^{-1}$]}
\tablenotetext{c}{The \citet{kel98} zodiacal light model gives a CIRB 4.0 kJy sr$^{-1}$ higher in the L band.}
\end{deluxetable}

\clearpage

\clearpage

\clearpage

\clearpage

\clearpage

\end{document}